\documentclass[floatfix,twocolumn,aps,prd,superscriptaddress,longbibliography]{revtex4-1}

\usepackage{graphicx}% Include figure files
\usepackage{dcolumn}% Align table columns on decimal point
\usepackage{bm}% bold math
\usepackage{textcomp}
\usepackage[intlimits]{amsmath}
\usepackage{esint}
\usepackage{braket}
\usepackage{color}
\usepackage{filecontents}
\usepackage{soul} %for \hl and \st

\soulregister\ref{7}
\soulregister\eqref{7}
\soulregister\cite{7}
\soulregister\onlinecite{7}

\begin{document}
	
	\preprint{APS/123-QED}
	
	\title{Transverse localization of transmission eigenchannels}% Force line breaks with \\
	
	\author{Hasan Y{\i}lmaz}
	\affiliation{Department of Applied Physics, Yale University, New Haven, Connecticut 06520, USA}%
	\author{Chia Wei Hsu}
	\affiliation{Department of Applied Physics, Yale University, New Haven, Connecticut 06520, USA}%
	\author{Alexey Yamilov}
	\affiliation{Department of Physics, Missouri University of Science \& Technology, Rolla, Missouri 65409, USA}%
	\author{Hui Cao}
	\email{hui.cao@yale.edu}
	\affiliation{Department of Applied Physics, Yale University, New Haven, Connecticut 06520, USA}%
	
	\date{\today}% It is always \today, today,
	%  but any date may be explicitly specified
	
	\begin{abstract}
	Transmission eigenchannels are building blocks of coherent wave transport in diffusive media, and selective excitation of individual eigenchannels can lead to diverse transport behavior. An essential yet poorly understood property is the transverse spatial profile of each eigenchannel, which is critical for coupling into and out of it. Here, we discover that the transmission eigenchannels of a disordered slab possess localized incident and outgoing profiles, even in the diffusive regime far from Anderson localization. Such transverse localization arises from a combination of reciprocity, local coupling of spatial modes, and nonlocal correlations of scattered waves. Experimentally, we observe signatures of such localization despite finite illumination area. Our results reveal the intrinsic characteristics of transmission eigenchannels in the open slab geometry, commonly used for applications in imaging and energy transfer through turbid media. 
	\end{abstract}
	
	\pacs{Valid PACS appear here}
	\maketitle
	
	Spatial inhomogeneities in the refractive index of a disordered medium cause multiple-scattering of light. 
	In disordered media such as biological tissue, white paint, and clouds, most of the incident light reflects back, hindering the transfer of energy and information through the media.
	However, by utilizing the interference of scattered waves, it is possible to prepare optimized wavefronts that completely suppress reflection---a striking phenomenon first predicted in the context of mesoscopic electron transport~\cite{Dorokhov, 1986_Imry_EPL, Mello1, Nazarov}.
	The required incident wavefronts are the eigenvectors of $t^{\dagger}t$ where $t$ is the field transmission matrix; the corresponding eigenvalues give the total transmission.  
	In a lossless diffusive medium, the transmission eigenvalues $\tau$ span from 0 to 1, leading to closed ($\tau \approx 0$) and open ($\tau \approx 1$) channels.
	In recent years, spatial light modulators (SLMs) have been used to excite the open channels~\cite{MoskR, VellekoopR, RotterR, Mosk2, Choi3, Choi4, Popoff2, 2016_Bosch_OE, Wade1, Cao4} to enhance light transmission through diffusive media. 
	Selective excitation of individual channels can dramatically change the total energy stored inside the random media as well as the spatial distribution of energy density~\cite{Choi1, Aubry, Genack1, Mosk3, Cao4, Cao5, 2017_Hong_arXiv}.
	
	However, some important questions regarding the transmission eigenchannels remain open. 
	What are the transverse spatial profiles for coupling light into such channels?
	Once coupled in, how do the eigenchannels spread in the transverse direction?
	In the Anderson localization regime of transport, a high-transmission channel is formed by coupled spatially localized modes~\cite{1987_Pendry_JPC, 2005_Bertolotti_PRL, 2006_Sebbah_PRL, Choi2, 2014_Pena_ncomms, Carminati1}; thus a transversely localized excitation and propagation is expected.
	However, Anderson localization is extremely hard to achieve in three-dimensional (3D) disordered systems~\cite{Page}, and diffusive transport is much more common. 
	In the diffusive regime, the open channels are expected to cover the entire transverse extent of the system~\cite{Choi1, Choi2}, utilizing all available spatial degrees of freedom.
	
	\begin{figure*}[th]
		\centering
		\includegraphics[width=\linewidth]{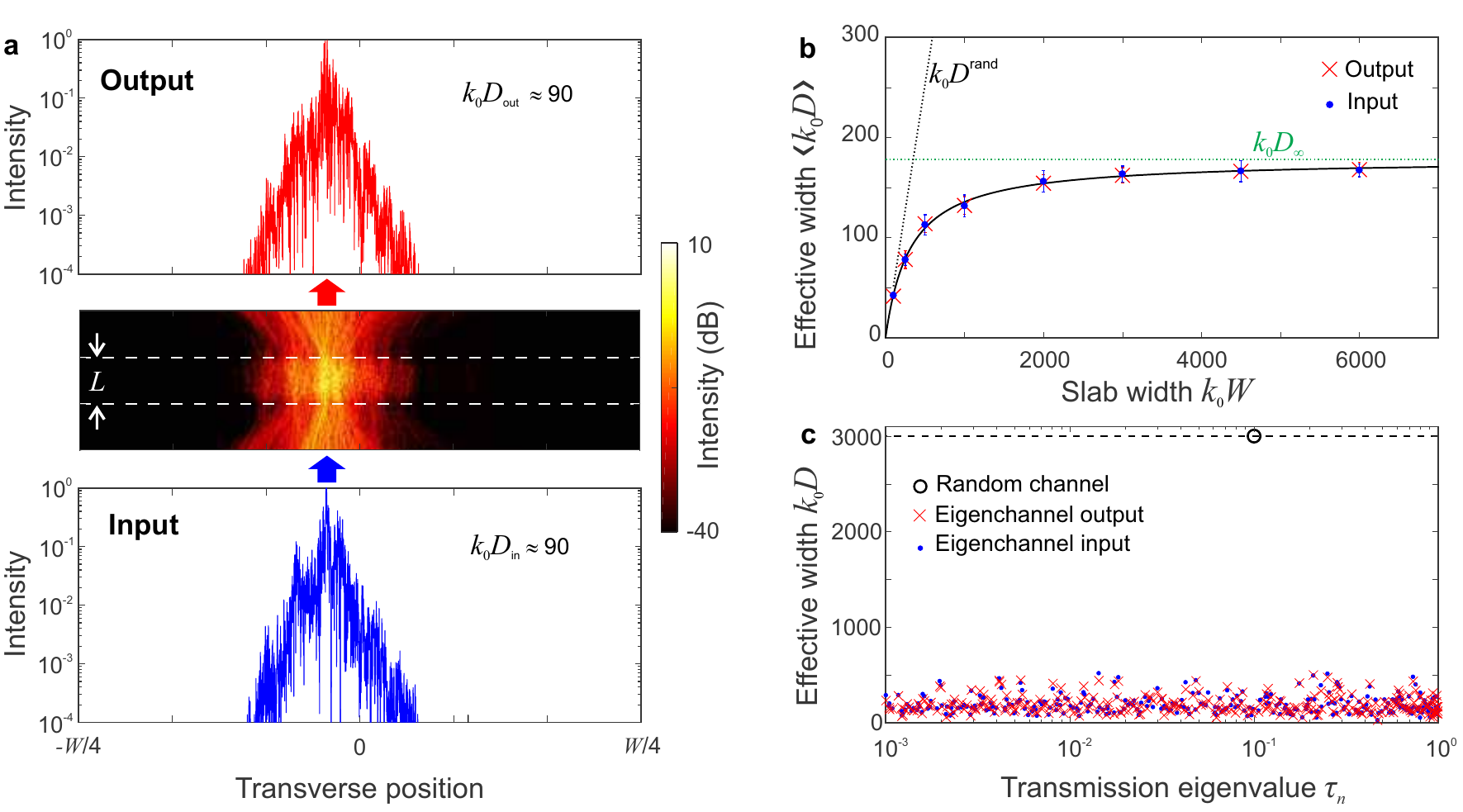}
		\caption{\textbf{Transverse localization of transmission eigenchannels.}
			\textbf{a}, Numerically calculated intensity profile of the highest-transmission eigenchannel through a two-dimensional (2D) diffusive slab, revealing localization in the transverse direction.
			The spatial distribution of field intensity is shown in the middle panel, with white dashed lines indicating boundaries of the slab; the relative vertical to horizontal scale is set to 5:1 for better visibility.
			The lower and upper panels show the incident and the transmitted intensity profiles on the input and output surfaces of the slab.
			This eigenchannel has a total transmission of $\tau_1 = 0.9999$, and its input width $D_{\rm in}$ and output width $D_{\rm out}$ are an order of magnitude smaller than the width $W$ of the slab. The normalized width of the 2D disordered slab is $k_0W = 6000$, the thickness $k_0L = 50$, the transport mean free path $n_0k_0l_t = 4.6$, the effective refractive index $n_0 = 1.5$, and the average transmission $\langle \tau \rangle = 0.10$.
			\textbf{b}, Input and output widths $D_\text{in}$ and $D_{\rm out}$ of open channels in disordered slabs of a constant thickness $k_0L = 50$ and a constant transport mean free path $n_0k_0l_t = 4.6$ but increasing widths $k_0W$, approaching an asymptotic value $D_{\infty}$ (green dashed line) in the wide-slab limit due to transverse localization.
			When $W$ is sufficiently smaller than $D_{\infty}$, the open channels fill the entire slab laterally and have widths close to the random incident wavefronts (black dashed line).
			Each data point is an average over the widths of all eigenchannels with $\tau_n \geq 1/\mathrm{e}$ in 10 realizations of structural disorder, with the error bars showing the standard deviation among realizations.
			The black solid line is the fitting that gives $D_{\infty}$ in the $W \to \infty$ limit.
			\textbf{c}, Eigenchannel widths versus transmission eigenvalues $\tau_n$ for the same diffusive slab as in part a, revealing that all eigenchannels are transversely localized with widths much smaller than those of random wavefronts.
		}
		\label{figure1}
	\end{figure*}
	
	Here we discover that the transmission eigenchannels are transversely localized even in the diffusive regime of transport.
	In a disordered slab of width $W$ much larger than thickness $L$, all transmission eigenchannels have a finite transverse extent that is much smaller than $W$.
	In the $W \to \infty$ limit, the channel width approaches an asymptotic value $D_\infty$, which scales as $(kl_t)L$ in two dimensions.
	Here $l_t$ is the transport mean free path, $k = n_0 k_0 = n_0 2\pi/\lambda$, $\lambda$ is the vacuum wavelength, and $n_0$ is the effective refractive index of the slab. 
	Furthermore, the eigenchannels do not spread laterally as they propagate through the slab, and the transverse extent at the output surface is equal to that at the input surface.
	Experimentally, we observe the transverse localization for high-transmission channels in a diffusive slab of zinc oxide (ZnO) nanoparticles. 
	The finite illumination area modifies the transmission eigenchannels, especially when the size of the illumination region is smaller than $D_\infty$. 
    While the lateral spreading is suppressed for high-transmission channels, it becomes enhanced for low-transmission channels. These properties can be explained in terms of optical reciprocity, bandedness of real-space transmission matrix, and non-local correlations of scattered waves as a result of multipath interference.
	The transverse localization of transmission eigenchannels that we discover in the diffusive regime is a distinct physical phenomenon from the previously known transverse localization of waves in Anderson-localized systems \cite{Lagendijk, Segev1, Mafi1, Wong, VanTiggelen, 2010_VanTiggelen_PRE}.
	The localization demonstrated here enables selective excitation of individual eigenchannels over input areas substantially smaller than the full extent of a diffusive slab, facilitating coherent control of light penetration and energy distribution in practical conditions. It therefore has potential impact on the advancement of deep-tissue imaging methods~\cite{YangR, ChoiR, 2015_Park_R} and the manipulation of light--matter interactions inside turbid media~\cite{Vynck1, Cao7}.

	\section{Transverse localization of eigenchannels}
	
	For a complete characterization of the transmission eigenchannels, we start with numerical simulations where we can exert full control over the incident wavefront and systematically explore the entire parameter space of interest.
	We solve the two-dimensional (2D) scalar wave equation $[\nabla^2 + k_0^2 \epsilon({\bf r})] \psi({\bf r}) = 0$ on a finite-difference grid.
	We consider disordered slabs of width $W$ and thickness $L$ in background refractive index $n_0$. The dielectric constant of the slab is modeled as $\epsilon({\bf r}) = n_0^2 + \delta \epsilon({\bf r})$ at each grid point, and $\delta \epsilon({\bf r})$ is a random number drawn from a zero-mean uniform distribution whose width determines the transport mean free paths $l_t$; see section B of the supplement for details.
	After calculating the field transmission matrix $t$ for the entire slab using the recursive Green's function method~\cite{1991_Baranger_PRB}, we obtain the incident wavefronts $\psi_n^{\rm in}$ of the eigenchannels via $t^\dagger t \psi_n^{\rm in} = \tau_n \psi_n^{\rm in}$, and calculate the spatial profile of the eigenchannels given such incident wavefronts.
	In this work we focus on scattering systems in the diffusive regime of transport, namely $Nl_t \gg L \gg l_t$, where $N \approx kW/\pi$ is the number of modes.
	
	Remarkably, we observe that in wide slabs, the eigenchannels are spatially localized in the transverse direction parallel to the slab; an exemplary open channel is shown in Fig.~\ref{figure1}a.
	Even though we impose no constraint on where or how wide the incident wavefront should be, the resulting open channel only occupies a relatively small transverse extent, utilizing just a fraction of the spatial degrees of freedom that are available across the width of the structure.
	Moreover, the open channel does not spread laterally as it propagates through the disordered slab; the transmitted profile is also localized, with a width similar to that of the incidence.
	As shown in the log-linear plot in Fig.~\ref{figure1}a, the transverse profile decays exponentially on both input and output surfaces, which is surprising given the wave transport is diffusive.
	
	A legitimate question is whether such transverse localization of eigenchannels persists in large systems, as experimentally the slab width $W$ is typically so large that it can be regarded infinite.
	To find the answer, we carry out a scaling analysis with increasing $W$, with results shown in Fig.~\ref{figure1}b.
	We quantify the width of an eigenchannel via the definition of participation number; the input diameter is found from the expression $D_{\rm in}\equiv\left[\int_{0}^{W}\mid\psi(x, z=0)\mid^2dx\right]^2 \Big/ \left[\int_{0}^{W}\mid\psi(x, z=0)\mid^4dx\right]$ where $\psi(x, z=0) $ is the field distribution of the channel at the input surface ($z=0$) of the slab, and similarly we calculate the output diameter $D_{\rm out}$ using the field distribution at the output surface ($z=L$).
	For each $W$, we consider all open channels (defined as having transmission eigenvalues $\tau_n \geq 1/\mathrm{e}$) in 10 different realizations of disorder.
	As shown in Fig.~\ref{figure1}b, we find $D_{\rm in}$ and $D_{\rm out}$ to be the same after ensemble average.
	In the $W \to \infty$ limit of interest, the open channel remains transversely localized, and its width saturates to an asymptotic value that we denote $D_{\infty}$.
	The extrapolation of $D_{\infty}$ in the $W\rightarrow\infty$ limit is described in the supplement section B.
	
	The absence of eigenchannel spreading, $\langle D_{\rm in} \rangle = \langle D_{\rm out} \rangle$, can be explained by reciprocity.
	Lorentz reciprocity requires the scattering matrix to be symmetric~\cite{2013_Jalas_nphoton}, so the transmission matrix coming from one side must be the transpose of the transmission matrix coming from the other side.
	One can express the transmission matrix through its singular value decomposition, $t = U \sqrt{\tau} V^\dagger$, where the $n$-th column of $V$ and $U$ are the normalized input and output wavefronts of the $n$-th transmission eigenchannel with eigenvalue $\tau_n$.
	Since $t^{\rm T} = V^* \sqrt{\tau} (U^*)^\dagger$, reciprocity demands that the phase conjugation of the $n$-th eigenchannel output must be precisely the input of the $n$-th eigenchannel coming from the other side, with the same eigenvalue.
	If the disordered medium is statistically equivalent for light incident from either side, the eigenchannel input width must be statistically identical for both directions. Thus the input and output channel widths should be the same after ensemble average.
	
	The above argument applies to all eigenchannels, open or closed.
	In fact, numerical simulations indicate that all eigenchannels are transversely localized with no lateral spreading, as shown in Fig.~\ref{figure1}c.
	In this example (same as in Fig.~\ref{figure1}a), the system width is $k_0W = 6000$, and random incident wavefronts have an average width of $k_0 D_{\rm in}^{\rm rand}=k_0W/2 = 3000$ from the participation number, but all eigenchannels have widths an order of magnitude smaller.
	For the closed channels, the transmitted intensities are much weaker than the incident ones, but the width of the transmitted profile (as defined by $D_{\rm out}$) remains the same as that of the incidence.

	\begin{figure}[th]
		\centering
		\includegraphics[width=\linewidth]{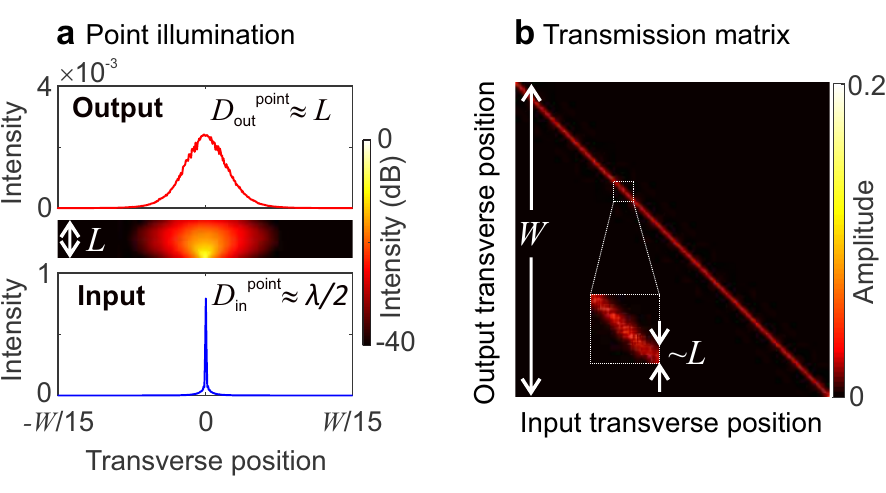}
		\caption{\textbf{Bandedness of real-space transmission matrix.}
			\textbf{a}, Calculated intensity profile inside a disordered slab when the incident light is focused to a diffraction-limited spot at the front surface, showing the extent of transverse spreading as light diffuses through the slab.
			$D_\mathrm{in}^\text{point}$ and $D_\mathrm{out}^\text{point}$ are the beam widths at the input and output surfaces.
			The intensity profiles shown are ensemble averaged over 1000 realizations of disorder. \textbf{b}, Amplitudes of the elements of the real-space transmission matrix.
			While the matrix size is given by the slab width $W$, only elements within a distance $\sim L$ to the diagonal are non-vanishing, because the extent of diffusive spreading in the slab is much less than the slab width.
			The simulation parameters are the same as in Fig.~1a. The inset is an expanded view of a part of the banded transmission matrix. 
		}
		\label{figure2}
	\end{figure}
	
	\begin{figure*}[th]
		\centering
		\includegraphics[width=\linewidth]{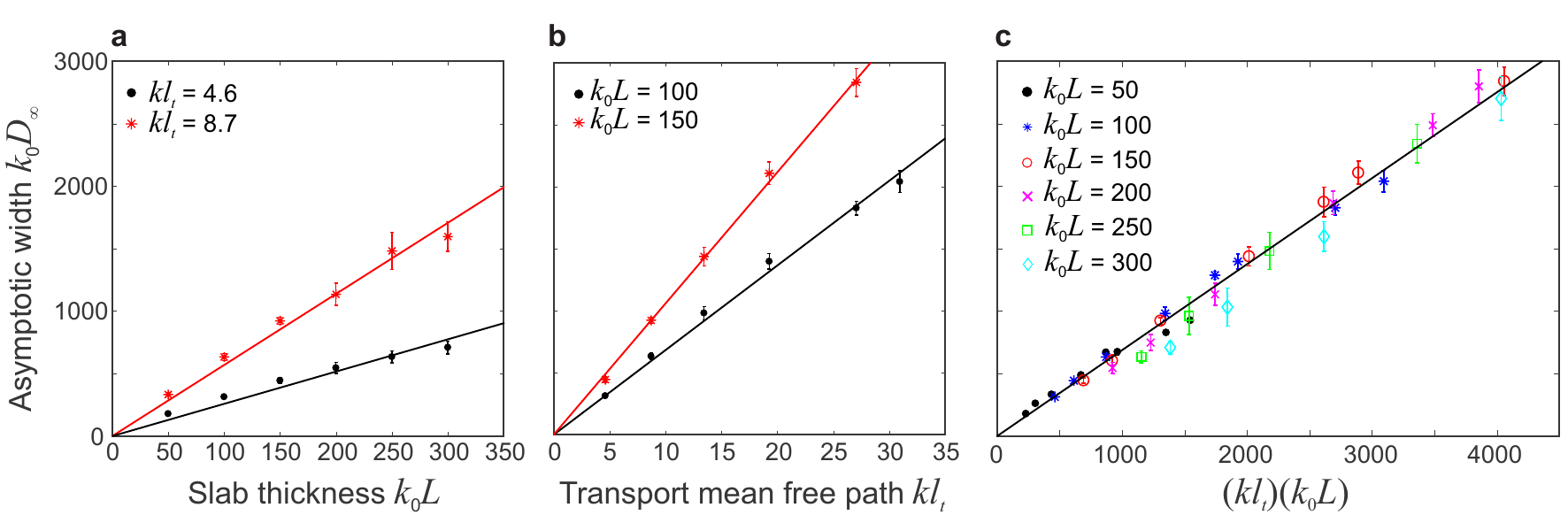}
		\caption{\textbf{Scaling of the asymptotic open channel width in diffusive slabs.} \textbf{a}, The asymptotic width $D_\infty$ of open channels as a function of the slab thickness $L$ when the transport mean free path $l_t$ is fixed. The solid lines represent a linear fit. The slope is smaller when $l_t$ is shorter.
		\textbf{b}, $D_\infty$ as a function of $l_t$ for fixed $L$. The solid lines are a linear fit, and the slope increases with increasing $L$.
		\textbf{c}, $D_\infty$ for diffusive slabs with different $L$, $l_t$, and $n_0$, showing a universal scaling $D_\infty \propto (n_0k_0l_t)L$. Linear regression gives the proportionality constant to be 0.68 (black solid line).
			Each data point represents an ensemble average as $\langle k_0D\rangle$ over open channels (channels that $\tau_n\geq 1/\text{e}$) and 10 realizations of disorder; the error bars are the standard deviation among 10 disorder realizations. Averaging as $1/\langle 1/k_0D\rangle$ gives the same scaling with respect to $l_t$ and $L$ (see Fig.~\ref{figure9} in the supplement).
		}
		\label{figure3}
	\end{figure*}
	
	\section{Origin of transverse localization}
	While reciprocity explains the absence of lateral spreading, it remains to be answered why the eigenchannels are transversely localized in the first place.
	We can gain insight by examining the real-space transmission matrix.
	Although scattering ensures that light with a specific incident angle is coupled into all outgoing angles once $L$ exceeds $l_t$, this is not the case in real space.
	Given a point-like excitation at the input surface, light spreads laterally as it diffuses through the disordered slab, covering a finite extent of width on the order of $L$ at the output surface; this is shown in Fig.~\ref{figure2}a.
	Such geometric local spreading is the origin of the much celebrated ``memory effect''~\cite{1988_Freund_PRL, 1989_Berkovits_PRB, 2015_Judkewitz_nphys, Vellekoop2}.
	As a result, the input and output spatial modes are not fully mixed, which emerges as non-vanishing elements only within a distance of $\sim L$ to the diagonal of the real-space transmission matrix ({\it i.e.}, the surface-to-surface Green's function), as shown in Fig.~\ref{figure2}{b}.
    It is noteworthy that 2D Anderson localization is absent in our systems, since the real-space transmission matrix bandwidth is proportional to the sample thickness in all of the systems we study here (see Fig.~\ref{figure8} in the supplement). Similarly, the real-space matrix $t^\dagger t$ also exhibits a bandwidth proportional to $L$.
	
	Random matrices with dominant near-diagonal elements were previously studied in the context of quantum chaos; it was found that the eigenvectors of such ``band random matrices'' are exponentially localized~\cite{Izrailev, IzrailevR, Mirlin}.
	It is therefore tempting to explain the transverse localization of eigenchannels through the ``bandedness'' of real-space transmission matrix for a wide slab.  
	The standard theory of band random matrices predicts that when the elements of a Hermitian random matrix is non-vanishing within a band of size $b$, the eigenvectors are localized with participation numbers proportional to $b^2$~\cite{Izrailev, IzrailevR, Mirlin}.
	In the present context, one would then expect the normalized eigenchannel width $kD$ to be on the order of $(kL)^2$ since the dimensionless bandwidth is $b \approx kL$.
	For the example in Fig.~\ref{figure1}, this argument suggests $k D_\infty \approx 5600$, but the actual eigenchannel width is only 90.
	The far smaller channel width indicates a much stronger transverse localization, which is beyond the standard band random matrix theory.
	
	To explore what determines the asymptotic open channel width $D_\infty$, we carry out a systematic study to map out its dependence on the slab thickness $L$ and the transport mean free path $l_t$.
	As shown in Fig.~\ref{figure3}a, the open channel width $D_\infty$ in fact scales {\it linearly} with the slab thickness $L$ that determines the real-space transmission matrix bandwidth $b$, in contrast to predictions from the standard band random matrix theory.
	Meanwhile, even though the transport mean free path $l_t$ does not affect the real-space transmission matrix bandwidth $b$, we find in Fig.~\ref{figure3}b that the open channel width $D_\infty$ also scales linearly with $l_t$.
	A dimensional analysis and the scale invariance of the electromagnetic wave equation indicates a prefactor proportional to the wave number $k = n_0 k_0$.
	Putting these together, we expect a scaling of $D_\infty \propto (kl_t)L$.
	In Fig.~\ref{figure3}c, we plot the compiled data of $D_\infty$ as a function of $(kl_t)L$ from $6 \times 6 = 36$ combinations of $(L, l_t)$ for $n_0 = 1.5$ and $6 \times 2 = 12$ combinations of $(L, l_t)$ for $n_0 = 1$; each $D_\infty$ is determined from 8 widths of $W$ and 10 realizations of disorder (totaling $>3000$ simulations).
	Indeed we observe the $D_\infty \propto (kl_t)L$ scaling.
	Least-square fit determines the proportionality constant to be 0.68, close to $2/3$.
	Therefore, we find the asymptotic open channel width $D_\infty \approx (2/3)(kl_t)L$ in 2D.
	Note that previous studies~\cite{Choi1, Choi2} did not find such transverse localization in the diffusive transport regime because the system width $W$ used in the previous simulations were not wide enough. Also note that such eigenchannel width $D_\infty$ is generally far smaller than the 2D localization length $\xi_{\rm 2D} \approx l_t e^{\pi k l_t/2}$.
	
	The reduction of eigenchannel width from $kL^2$ to $kl_tL$ requires explanations beyond the bandedness of the real-space transmission matrix.
	The key factor is the correlations among the non-zero matrix elements induced by multiple scattering of light in the slab. 
	It is known that multipath interference in scattering media leads to non-local correlations of scattered waves~\cite{Cwilich, Stone2, 1988_Mello_PRL, Shapiro, BerkovitsPR94, Genack0, Scheffold1, 2002_Sebbah_PRL, Yamilov, Muskens1, Carminati2, 2018_Bertolotti_PRX,Akkermans}. When we replace the non-vanishing elements of the real-space transmission matrix with uncorrelated complex Gaussian random numbers ({\it i.e.} remove the correlations artificially), we observe much wider eigenchannel widths that scale as $kL^2$ as predicted by standard band random matrix theory. Stronger scattering (smaller $kl_t$) enhances non-local correlations and leads to tighter transverse localization of the eigenchannels.
    
	Extending such scaling study to disordered slabs in 3D is a daunting computational task. Nevertheless, we expect transverse localization of transmission eigenchannels in 3D both at the input and the output surfaces, since 3D systems also possess banded real-space transmission matrices, non-local correlations, and reciprocity.
	
	\begin{figure*}[ht]
		\centering
		\includegraphics[width=\linewidth]{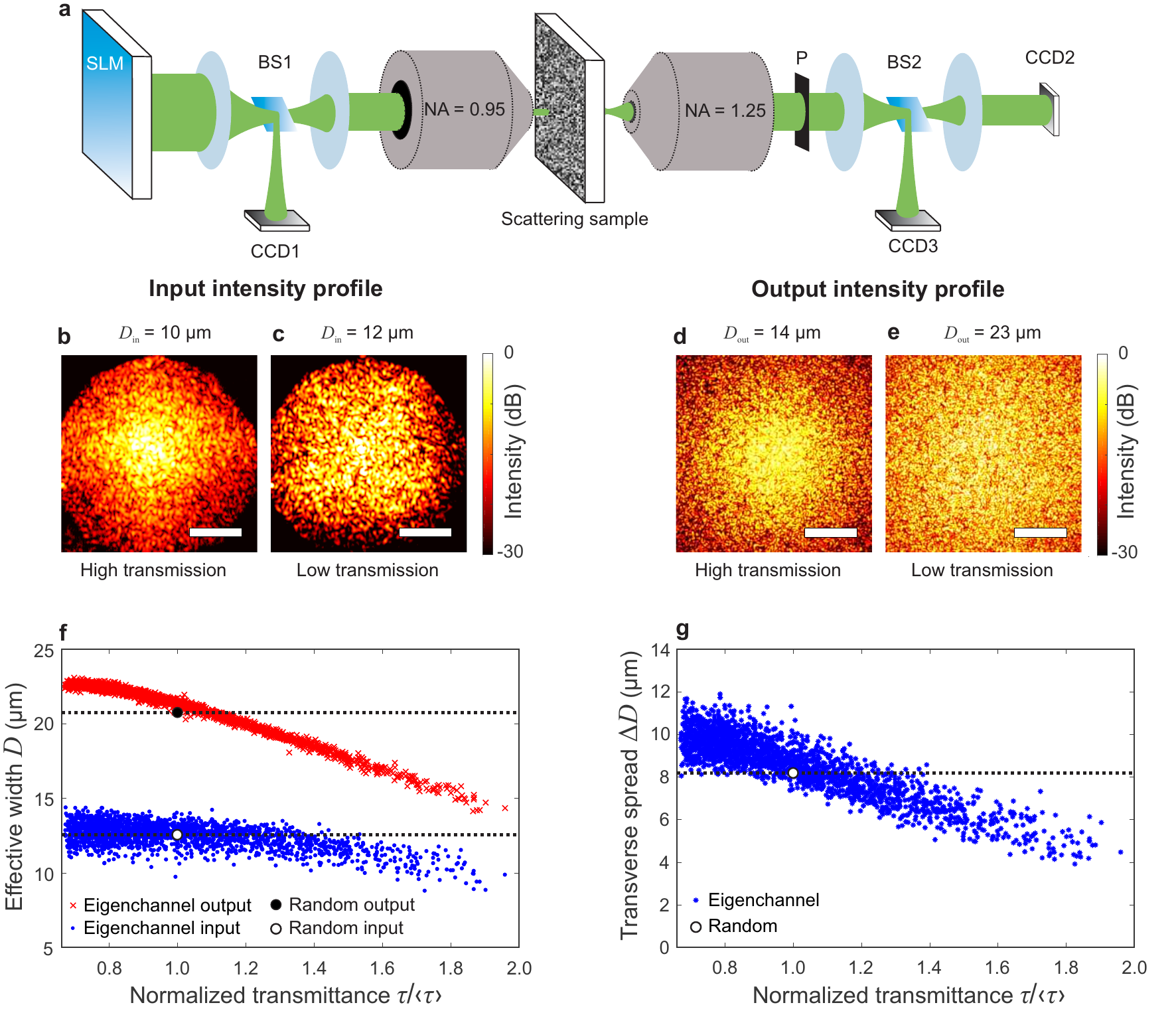}
		\caption{\textbf{Experimental evidence of transverse localization for high-transmission channels.} 
			\textbf{a}, A simplified schematic of the experimental setup for measuring the field transmission matrix of a three-dimensional (3D) disordered slab with a finite illumination area, followed by selective excitation of individual transmission eigenchannels and measurement of their intensity profiles on input and output surfaces of the sample. SLM: spatial light modulator, BS: beam splitter, CCD: charge-coupled device camera, NA: numerical aperture, P: linear polarizer. 
			\textbf{b,c}, Measured intensity profile of the incident wavefront on the front surface of the sample, for a high-transmission (b) and a low-transmission (c) eigenchannel.
			\textbf{d,e}, Corresponding intensity profiles on the back surface of the sample.
			Scale bars: 6 \textmu m.
			\textbf{f}, Input width (blue filled circles) and output width (red crosses) of all 2048 eigenchannels as a function of the normalized transmission eigenvalue $\tau/\langle\tau\rangle$. The dashed lines denote the input width (black open circle) and output width (black filled circles) for random incident wavefronts.  
			\textbf{g}, Transverse spreading length $\Delta D = D_\mathrm{out} - D_\mathrm{in}$ versus the normalized transmission eigenvalue $\tau/\langle\tau\rangle$ for experimentally measured transmission eigenchannels; (see Fig.~\ref{figure7} in the supplement for predicted and measured values of $\tau/\langle\tau\rangle$).
			Compared to random wavefronts, the transverse spreading is suppressed for high-transmission channels, but enhanced for low-transmission channels.
		}
		\label{figure4}
	\end{figure*} 
	
	\section{Experimental results}
		
	To search for experimental evidence of transverse localization of transmission eigenchannels, we measure the spatial profiles of individual eigenchannels at the input and output surfaces of a 3D scattering slab. The sample consists of zinc oxide (ZnO) nanoparticles that are spin-coated on a cover slide. The thickness of the ZnO layer is about 10 \textmu m, much less than the lateral width of the layer (2 cm $\times$ 2 cm). The average transmittance of light at wavelength of 532 nm through the sample is approximately 0.2.

	We start by measuring the transmission matrix of the disordered slab. A simplified schematic of the experimental setup is shown in Fig.~\ref{figure4}a, with a detailed one given in Fig.~\ref{figure6} in the supplement section A. A spatially uniform monochromatic laser beam at wavelength $\lambda = 532$ nm is modulated by a phase-only SLM. The SLM surface is imaged by a pair of lenses onto the pupil of a microscope objective. Therefore, the spatial profile of illumination is the 2D Fourier transform of the SLM phase pattern; the illumination area is finite, and its widths scales inversely with the SLM macropixel size. We use the SLM and a CCD camera to measure the field transmission matrix in $k$-space, using a common-path interferometry method akin to references~\cite{Popoff1, Wade1}. The number of SLM macro-pixels that modulate the input beam is 2048, and the number of output speckle grains recorded by the camera is about 15000.

	After measuring the field transmission matrix $t$, we determine the incident wavefronts of individual eigenchannels as the eigenvectors of $t^\dagger t$. Then we display the corresponding phase patterns on the SLM, and record the 2D spatial intensity profiles $I(x,y)$ at the input and output surfaces of the sample with two cameras (CCD1, CCD3). We define the effective area of such a profile through the 2D participation number $A \equiv \left[\iint I(x,y) \mathrm{d}x \, \mathrm{d}y  \right]^2 / \left[\iint I^2(x,y) \mathrm{d}x \, \mathrm{d}y\right]$, and the effective width $D$ through $A = \pi \left(D/2\right)^2$. The highest-transmission eigenchannel indeed exhibits narrower spatial profiles (shown in Figs.~\ref{figure4}b,d): it has $D_\mathrm{in} \approx 10$ \textmu m and $D_\mathrm{out} \approx 14$ \textmu m, while random wavefronts have $D_{\rm in}^{\rm rand} \approx 13$ \textmu m and $D_{\rm out}^{\rm rand} \approx 21$ \textmu m.
% * <yilmazhasan@gmail.com> 2018-06-02T15:08:58.224Z:
%
% > $A \equiv \left[\iint I(x,y) \mathrm{d}x \, \mathrm{d}y  \right]^2 / \left[\iint I^2(x,y) \mathrm{d}x \, \mathrm{d}y\right]$,
%
% ^.
The lateral spreading is also less: $\Delta D = D_\mathrm{out} - D_\mathrm{in} \approx 4$ \textmu m for the highest-transmission eigenchannel, while $\Delta D^{\rm rand} \approx 8$ \textmu m $\sim L$ for random wavefronts.
These are experimental signatures of the transverse localization phenomenon in high-transmission eigenchannels introduced in the previous section.

	More complex behaviors emerge when we also examine eigenchannels with lower transmission. In our experiment, the low-transmission eigenchannels have incident profiles (Fig.~\ref{figure4}c) comparable in size to those of random wavefronts. Furthermore, their output profiles (Fig.~\ref{figure4}e) are wider than output profile of a random input in terms of the participation number $D_\mathrm{out}$. Figs.~\ref{figure4}f,g show the width and spreading of all of the 2048 eigenchannels as a function of the normalized transmission eigenvalue, and compare them to random incident wavefronts. We can see that, in contrast to Fig.~\ref{figure1}c, the experimental eigenchannel widths reveal systematic dependences on the transmission eigenvalue, particularly for $D_\mathrm{out}$. Specifically, the transverse spreading increases with decreasing eigenvalue, with the high-transmission eigenchannels exhibiting suppressed lateral spreading and the low-transmission eigenchannels exhibiting enhanced spreading. In next section, we demonstrate that this discrepancy between numerical and experimental results can be understood by taking into account the finite illumination area, phase-only modulation, and the noise in the experiment.

	\section{Effect of incomplete control}
    
        \begin{figure*}[ht]
		\centering
		\includegraphics[width=\linewidth]{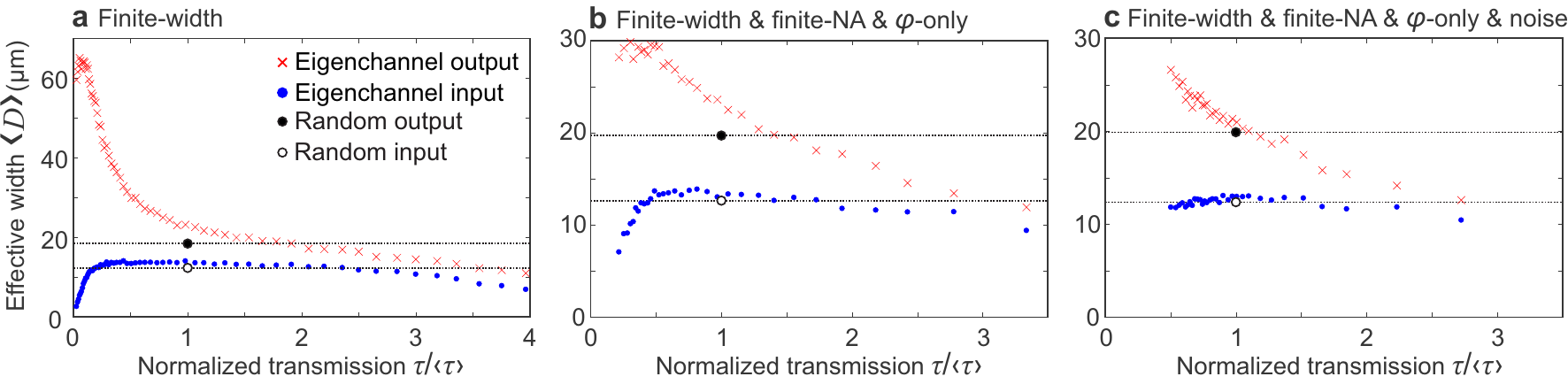}
		\caption{\textbf{Modification of transmission eigenchannel widths by incomplete control}. \textbf{a}, Numerically calculated input width (blue filled circles) and output width (red crosses) of all transmission eigenchannels as a function of the normalized transmission eigenvalue in a 2D slab with local illumination. Each data point is ensemble averaged over 50 realizations of disorder. For random incident wavefront, the beam width at the front surface of the slab is $D_{\rm in}^{\rm rand} \approx 13$ \textmu m (black open circle), and the transmitted beam width at the back surface is $D_\text{out}^\text{rand} \approx 19$ \textmu m (filled black circle). \textbf{b}, Including finite numerical aperture of illumination and detection, as well as phase-only modulation of incident wavefront reduces the range of eigechannel widths. \textbf{c}, Adding random Gaussian noise to the transmission matrix further modifies the eigenchannel widths, especially for the low-transmission channels. The slab width is $W = 508$ \textmu m, the thickness is $L = 10$ \textmu m, the transport mean free path is $l_t = 1$ \textmu m, and the average refractive index is $n_0 = 1.4$. The slab is sandwiched between air (refractive index 1) and glass (refractive index 1.5). See Fig.~\ref{figure10} in the supplement for transverse spread $\Delta D$.}
		\label{figure5}
		\end{figure*}

	There are important differences between the experimental setup and the idealized scenario considered in Figs.~1--3. In our experiment, the illumination beam-width on the sample surface is finite and is comparable to $L$, much smaller than the expected $D_{\infty}$. Also, we use phase-only modulation over a finite fraction of incident angles, and collect a finite fraction of outgoing angles in one polarization. Such experimental conditions lead to incomplete control which is known to affect the transmission eigenvalues~\cite{Stone1, Wade1}, and we expect them to also modify the eigenchannel profiles. Experimentally it is not possible to separate the different factors, but we can do so with simulations. Numerically we consider 2D disordered slabs with parameters comparable to the experiment (see the caption of Fig.~\ref{figure5}), with the asymptotic open-channel width being $D_\infty \approx 90$ \textmu m. Naturally we do not expect quantitative comparison with the 3D sample in the experiment, but we aim to extract physical insights that do not depend on dimensionality.

	We describe finite-width illumination by grouping incident modes into equally-spaced intervals of transverse momenta that model the SLM macropixels~\cite{Wade1}. For random incident wavefronts, the beam widths as defined by the participation number are $D_{\rm in}^{\rm rand} \approx 13$ \textmu m on the front and $D_\text{out}^\text{rand} \approx 19$ \textmu m on the back surface.
With such finite-width illumination (Fig.~\ref{figure5}a), we find that eigenchannels with intermediate eigenvalues have incident widths $D_\text{in} \approx D_{\rm in}^{\rm rand}$; this is to be contrasted with the full-width illumination case of Fig.~\ref{figure1}c.
Meanwhile, it is striking that despite the illumination width $D_{\rm in}^{\rm rand}$ is much smaller than the asymptotic eigenchannel width $D_\infty$, both high-transmission and low-transmission channels have input widths even smaller than $D_{\rm in}^{\rm rand}$ (Fig.~\ref{figure5}a). We attribute this to the fact that these channels utilize multipath interference to enhance or suppress the total transmission. 
Indeed, in the scattering paths picture of wave propagation in disordered media, path crossings inside the sample lead to non-local correlations~\cite{BerkovitsPR94,Akkermans} and enhances the range of transmission eigenvalues~\cite{Wade1}. Therefore, eigenchannels with extremal eigenvalues prefer smaller input beam widths to increase the probability of crossing.
	In addition, the extremal eigenchannels preferentially enhance or suppress the output intensity near the center of the beam (see Fig.~\ref{figure11} in the supplement). Such a non-uniform modification of the transmitted intensity profile results in an effective reduction of the participation number $D_{\rm out}$ for the high-transmission eigenchannels that we observe in Fig.~\ref{figure5}a, and similarly for the increased $D_{\rm out}$ of the low-transmission eigenchannels. We find that the other sources of incomplete control have relatively minor effects. In Fig.~\ref{figure5}b, we include the phase-only modulation of the incident wavefront, as well as the finite range of numerical aperture (NA) both in illumination and detection (see section B of the supplement for detail). The ranges of transmission eigenvalues and eigenchannel widths both decrease, but the qualitative trends remain the same.

	Finally, we also model the effect of experimental noise (see section B of the supplement for detail). As shown in Fig.~\ref{figure5}c, the low-transmission eigenchannels are more sensitive to noise than the high-transmission channels: the input widths of low-transmission channels become equal to those of random incident wavefronts, while the input widths of the high-transmission channels only change slightly. These results agree qualitatively with our experimental data.
    
	\section{Conclusion}
	
	In conclusion, we discover transverse localization of transmission eigenchannels in diffusive slabs. In the presence of complete control, each eigenchannel has statistically identical input and output widths as a result of optical reciprocity. In a 2D slab, the asymptotic width for open channels is $D_\infty \approx (2/3)kl_tL$, due to the bandedness and non-local correlations of the real-space transmission matrix. We experimentally observe signatures of transverse localization of transmission eigenchannels in a diffusive slab with finite illumination area. While the transverse spreading is suppressed for high-transmission channels, it is enhanced for low-transmission channels. These results are reproduced numerically and explained via multipath interference effects. Our results provide physical insights of transmission eigenchannels in open slab geometry and illustrate the effects of local illumination on eigenchannel profiles. This work opens the possibility of controlling transmission eigenchannels in open systems, which will have significant impact on information and energy delivery through strongly scattering systems. 
    
    \section*{Acknowledgement}
We thank Allard Mosk, Azriel Genack, Boris Shapiro, Frank Scheffold, Sergey Skipetrov, Stefan Bittner, Stefan Rotter, and Tsampikos Kottos for stimulating discussions and useful feedback.   

This work was supported by the Office of Naval Research (ONR) under grant no. MURI N00014-13-0649, and by the US-Israel Binational Science Foundation (BSF) under grant no. 2015509.  

\section*{Author contributions}
H.Y. performed the experiments and analyzed the data. C.W.H. performed the numerical simulations and fabricated the samples. H.Y. analyzed the numerical data. C.W.H. helped with experimental data acquisition and contributed to numerical data analysis. H.C. supervised the project. All authors contributed to the interpretation of the results. H.Y. and C.W.H. prepared the manuscript, H.C. edited it, and A. Y. provided feedback.
    
   \section{Supplementary material}
    
    	This document provides supplementary information to ``Transverse localization of transmission eigenchannels". In the first section we describe details of our experimental setup and measurement procedure. In the second section, we present details of our numerical simulations.
        
        \subsection{Experiment}
	
	\begin{figure*}[htpb]
		\centering
		\includegraphics[width=\linewidth]{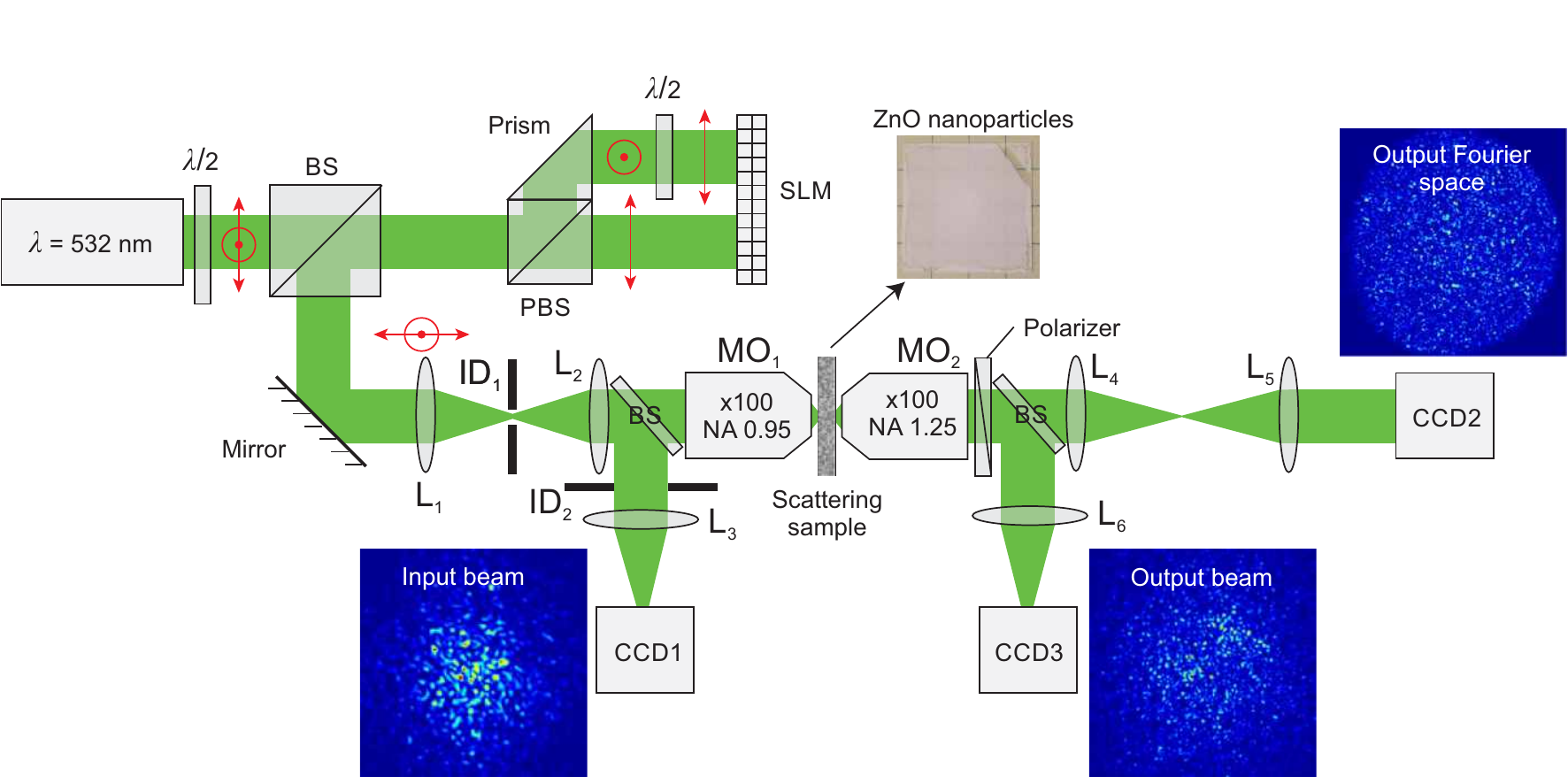}
		\caption{\textbf{Detailed experimental setup.} A reflective phase-only spatial light modulator (SLM) modulates the phase-fronts of orthogonal linear polarization components of a monochromatic laser beam ($\lambda = 532$ nm). The optical field transmission matrix of the scattering sample is measured in $k$-space with the SLM and the camera CCD2. The scattering sample, shown in the inset, is a 10 \textmu m thick film of ZnO nanoparticles with average transmittance 0.2, deposited on a glass substrate. Cameras CCD1 and CCD3 image the spatial intensity profiles of light on the front (input) and back (output) surfaces of the sample, respectively. $\lambda/2$, half-wave plate; BS, beam splitter; PBS, polarizing beam splitter; MO$_{1-2}$, microscope objectives; $L_{1-6}$, lenses; ${\rm ID}_{1-2}$: iris diaphragms.}
		\label{figure6}
	\end{figure*}

	The scattering sample in our experiment is made of closely-packed zinc oxide (ZnO) nanoparticles (average diameter $\sim$ 200 nm), deposited on a cover slip of thickness 170 \textmu m. The ZnO layer thickness is about 10 \textmu m, and the transport mean free path is approximately 1.5 \textmu m. The average transmission is approximately 0.2. We define the interface between ZnO and air as the front (input) surface, and the interface between the ZnO and the cover slip as the back (output) surface of the sample. The effective index of refraction for the ZnO nanoparticle layer is about 1.4, which almost matches the refractive index of the glass substrate (cover slip). 
		
	Our experimental setup is sketched in Fig.~\ref{figure6}. A linearly-polarized monochromatic laser beam (Coherent, Compass 215M-50 SL) with wavelength  $\lambda = 532$ nm is expanded and then clipped in order to uniformly cover a large area on the spatial light modulator (SLM). Its polarization direction is rotated from vertical to 45$^{\circ}$ by a half-wave $(\lambda/2)$ plate, and consequently split into vertical and horizontal polarizations by a polarizing beam splitter (PBS). The horizontal-polarized component of the beam illuminates one part of a reflective phase-only SLM (Hamamatsu, X10468-01). Since the SLM only modulates horizontal polarization, the vertical-polarized component of the beam is converted to horizontal polarization by another $\lambda/2$ plate before impinging onto the second part the SLM; the modulated reflected beam is converted back to vertical polarization after passing through the same $\lambda/2$ plate again. The two polarizations are recombined at the PBS, and the SLM plane is imaged onto the pupil of a microscope objective $\text{MO}_1$ (Nikon CF Plan 100$\times$ with a numerical aperture $\text{NA}_\mathrm{in} = 0.95$) by a pair of lenses $L_1$ and $L_2$ (with focal lengths $f_1 = f_2 = 200$ mm). This setup enables independent modulation of the spatial wavefront for two orthogonal polarizations. An iris diaphragm ${\rm ID}_1$ between $L_1$ and $L_2$ blocks high-order diffractions from the SLM. The objective $\text{MO}_1$ projects the Fourier transform of the SLM phase pattern onto the front (input) surface of the scattering sample. 
	
	To measure the spatial profile of illumination on the sample surface, we insert a beam splitter before the objective $\text{MO}_1$ to split the input beam, and use another lens $L_3$ ($f_3 = 100$ mm) to image the Fourier plane of the SLM onto a CCD camera CCD1 (Allied Vision, Guppy PRO F-031B). An iris ${\rm ID}_2$ between the beam splitter and $L_3$ blocks light that does not enter $\text{MO}_1$.
	
	In transmission, the Fourier transform of the transmitted field on the back (output) surface of the sample is imaged onto a CCD camera CCD2 (Allied Vision, Manta G-031B) by an oil-immersion microscope objective $\text{MO}_2$ (Edmund DIN Achromatic 100$\times$, $\text{NA}_\mathrm{out} = 1.25$) and a pair of lens $L_4$ ($f_4 = 200$ mm) and $L_5$ ($f_5 = 100$ mm). A linear polarizer is placed right after $\text{MO}_2$ to filter out one polarization component of the transmitted light. In between the polarizer and the lens $L_4$, a beam splitter is inserted to split the output beam, and the intensity profile on the back (output) surface of the sample is imaged onto another CCD camera CCD3 (Allied Vision, Pike F-100B) by a lens $L_6$ ($f_6 = 200$ mm).
	
	Field transmission matrix from the SLM to CCD2 is measured with a common-path interferometry akin to the method in reference \cite{Popoff1}. We display a complete set of 2048 input field vectors in Hadamard basis on 2048 macropixels of the SLM, each consisting of $4\times 4$ SLM pixels. The 2048 macropixels are imaged onto approximately half of the area of $\text{MO}_1$ pupil as the signal field. To measure phases of each Hadamard base vector, we display a random (but fixed) phase pattern on the remaining SLM macropixels (of the same $4\times 4$ size) that also get imaged onto $\text{MO}_1$ as reference macropixels. Together, the signal macropixels and the reference macropixels fill the the pupil of $\text{MO}_1$. In order to measure intensity of each Hadamard basis vector in transmission, a high-spatial-frequency phase grating is displayed on the reference region of the SLM so that light incident upon the reference macropixels is blocked by the iris ${\rm ID}_1$ and does not enter $\text{MO}_1$.
	
\begin{figure}[htpb]
		\centering
		\includegraphics[width=\linewidth]{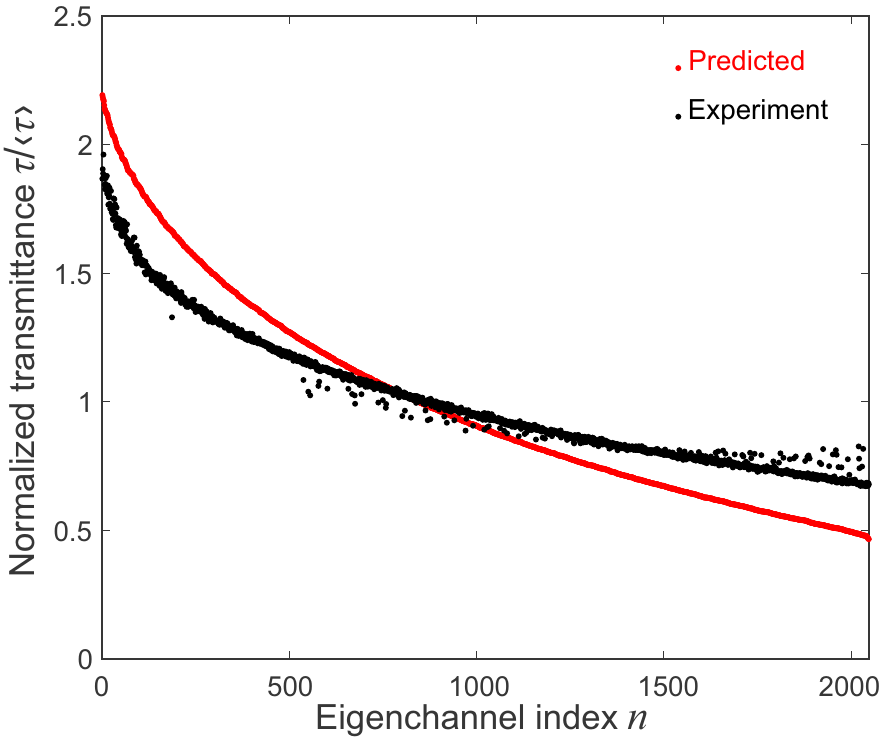}
		\caption{\textbf{Normalized transmission eigenvalues $\tau/\langle\tau\rangle$ for all eigenchannels.} Red filled circles represent values of $\tau/\langle\tau\rangle$ predicted from the measured transmission matrix after removing amplitude modulation from input eigenvectors. Black filled circles denote the experimentally measured $\tau/\langle\tau\rangle$ when the phases of input eigenvectors are displayed on the SLM. The latter has a smaller range than the former due to measurement noise in our experiment.}
		\label{figure7}
	\end{figure}

	After measuring the field transmission matrix, we calculate the eigenvectors which represent the input wavefronts for individual transmission eigenchannels:
	\begin{equation}
	\label{eigen}
	\tilde{t}^{\dagger}\tilde{t}\ket{\tilde{\psi}_n} = \tau_n\ket{\tilde{\psi}_n} \, ,
	\end{equation}
	where $\ket{\tilde{\psi}_n}$ is the $n$-th eigenvector, and $\tau_n$ is the corresponding eigenvalue that gives the transmittance of the $n$-th eigenchannel. After finding the eigenvectors, we display the phase pattern of $\ket{\tilde{\psi}_n}$ on the 2048 macropixels of the SLM, and record the intensity profiles on the front and back surfaces of the scattering sample. At this time, the high-spatial-frequency phase grating is displayed on the reference region of the SLM. Fig.~\ref{figure7} shows the normalized transmittance $\tau/\langle\tau\rangle$ of each eigenchannel in our experiment. The red filled circles denote the values of $\tau/\langle\tau\rangle$ predicted from the measured transmission matrix, and black filled circles represent the experimentally measured $\tau/\langle\tau\rangle$. While the range of $\tau/\langle\tau\rangle$ is predicted to be between 2.2 and 0.47, the experimental values of $\tau/\langle\tau\rangle$ range from 1.95 to 0.67 due to measurement noise.
    
	\subsection{Numerical simulations}

	In this section, we present details of our numerical simulations. The first subsection describes the extraction of transport and scattering mean free paths of the simulation samples. The second subsection depicts the relation between sample thickness $L$ and real-space transmission matrix bandwidth $b$. The third subsection narrates how the asymptotic eigenchannel widths are calculated for diffusive media in wide slab geometry. The final subsection provides details of numerical simulations that account for finite-width illumination, finite numerical aperture, phase-only control, and noise in the experiment.
  
    We simulate wave propagation through two-dimensional (2D) diffusive slabs numerically. The normalized width of a slab is $k_0W$ and the normalized thickness is $k_0L$, where $k_0 = 2\pi/\lambda$ and $\lambda$ is the vacuum wavelength. The slab is discretized on a  2D square grid, and the grid size is $(\lambda/2\pi)\times (\lambda/2\pi)$. The dielectric constant at each grid point is $\epsilon(\textbf{r}) = n_0^2 + \delta\epsilon(\textbf{r})$, where $n_0$ is the average refractive index of the disordered slab, $\delta\epsilon(\textbf{r})$ a random number drawn from the interval $[-\sigma,\sigma]$ with uniform probability. The disordered slab is sandwiched between two homogeneous materials with refractive indices of $n_1$ and $n_2$. Either perfectly reflecting (Dirichlet) or periodic boundary conditions are applied to the transverse boundaries.

	To obtain the field transmission matrix $t$ at wavelength $\lambda$, we solve the scalar wave equation $\left[\nabla^2+k_0^2\epsilon(\textbf{r})\right]\psi(\textbf{r}) = 0$ with the recursive Green's function method \cite{Stone3}. The singular value decompostion of $t = U \sqrt{\tau} V^\dagger$ gives the transmittance $\tau$, the input $V$ and output $U$ wavefronts of transmission eigenchannels.
    
        \subsubsection{Extraction of transport and scattering mean free paths}
    
    In this and the following two subsections, we set $n_1 = n_2 = n_0$ and apply the perfectly reflecting (Dirichlet) boundary conditions to the transverse boundaries ($n_1 = n_2 = n_0 = 1.5$ in this subsection). The transport mean free path $l_t$ is obtained from the average transmittance $\langle\tau\rangle$:
		\begin{equation}
		\label{tmfp}
		\langle\tau\rangle = \frac{\left(1+\Delta\right)l_t}{L+2\Delta l_t} \, ,
		\end{equation}
where $\Delta = 0.818$ for the 2D diffusive slab with index-matched homogeneous media at both surfaces \cite{Durian, Yamilov}.

    To extract scattering mean free path $l_s$ for a fixed strength of disorder $\sigma$, we calculate the transmission matrices of waveguides with the same width $k_0W = 100$ and varying thickness between $k_0L = 1$ and $k_0L = 6k_0l_t$. After average over 40000 disorder realizations, the diagonal elements of the transmission matrices $\langle\psi_{nn}\rangle$ give the scattering mean free path \cite{Genack6}
    	\begin{equation}
		\label{tmfp}
		\lvert \langle\psi_{nn}\rangle\rvert^2 = \mathrm{exp}\left[\frac{-kL}{k_z^nl_s}\right ] \ ,
		\end{equation}
where $k_z^n$ the longitudinal component of the wave vector for each waveguide mode $n$. Table~\ref{table} presents the values of the transport mean free path and the scattering mean free path for different strengths of disorder $\sigma$. Their values are approximately equal, $l_t\approx l_s$.

\begin{table}[htbp]
\centering
\caption{\bf Transport mean free path $l_t$ and scattering mean free path $l_s$ for each disorder strength $\sigma$.}
\begin{tabular}{c c c c c c c}
\hline
$\sigma$ & \ \ 0.60 \ \ & \ \ 0.65 \ \  & \ \ 0.75 \ \  & \ \ 0.90 \ \  & \ \ 1.10 \ \  & \ \ 1.45 \\
\hline                                                                                  
$kl_t$ & 30.9 & 27.1 & 19.3 & 13.5 & 8.7 & 4.6 \\                                 
$kl_s$ & 34.3 & 29.3 & 22.2 & 15.5 & 10.4 & 6.0 \\

\hline
\end{tabular}
\label{table}
\end{table} 

    \subsubsection{Bandwidth $b$ of real-space transmission matrices}
    
        \begin{figure}[ht]
		\centering
		\includegraphics[width=\linewidth]{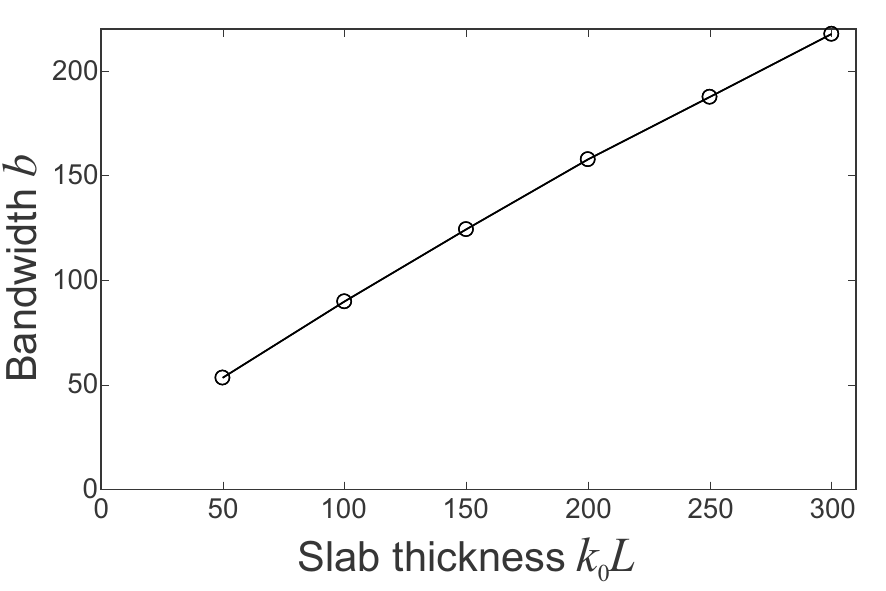}
		\caption{\textbf{Bandwidth $b$ of real-space transmission matrices versus the slab thickness $k_0L$.} The slab width is $k_0W = 6000$. The mean free path is the shortest among all simulated slabs. The linear scaling of bandwidth $b$ with the slab thickness $L$ confirms diffusive transport.}
		\label{figure8}
		\end{figure}
    
   We compute the bandwidth $b$ of real-space transmission matrices using the definition of the participation number
   \begin{equation} b\equiv\Big\langle\left[\int_{0}^{W}\mid\psi\mid^4dx\right]\Big/\left[\int_{0}^{W}\mid\psi\mid^2dx\right]^2\Big\rangle^{-1},
   \end{equation} 
   where $\langle\rangle$ is an ensemble average over all columns of ten real-space transmission matrices representing different realization of disorder. We observe the bandwidth $b$ of the real-space transmission matrix  scales linearly with the slab thickness $L$ within the range of disorder strength in our simulation.   Fig.~\ref{figure8} is a plot of $b$ versus $k_0L$ for the slabs with the shortest transport mean free path $kl_t = 4.6$. The linear scaling of $b$ with $L$ is an evidence that 2D Anderson localization is absent even in the slabs with the smallest $kl_t = 4.6$.

	\subsubsection{Asymptotic width of open channels}

        \begin{figure}[ht]
		\centering
		\includegraphics[width=\linewidth]{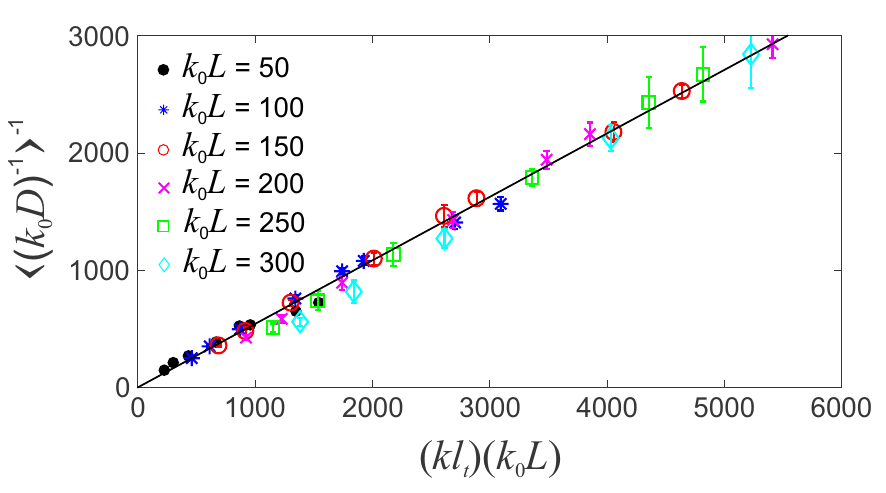}
		\caption{\textbf{Scaling of asymptotic width $D_\infty$ of transmission eigenchannels.} $k_0D_\infty$ is obtained from $ \langle \left(k_0D\right)^{-1}\rangle^{-1}$ in the limit $W\rightarrow \infty$. $D_\infty$ of diffusive slabs with different $L$, $l_t$, and $n_0$, showing a universal scaling $D_\infty \propto (n_0k_0l_t)L$. Linear regression gives the proportionality constant to be 0.54 (black solid line). Each data point represents an ensemble average over open channels (with $\tau_n\geq 1/\text{e}$) in ten realizations of disorder; the error bars are the standard deviation among the disorder realizations.}
		\label{figure9}
		\end{figure}
        
         \begin{figure*}[ht]
		\centering
		\includegraphics[width=\linewidth]{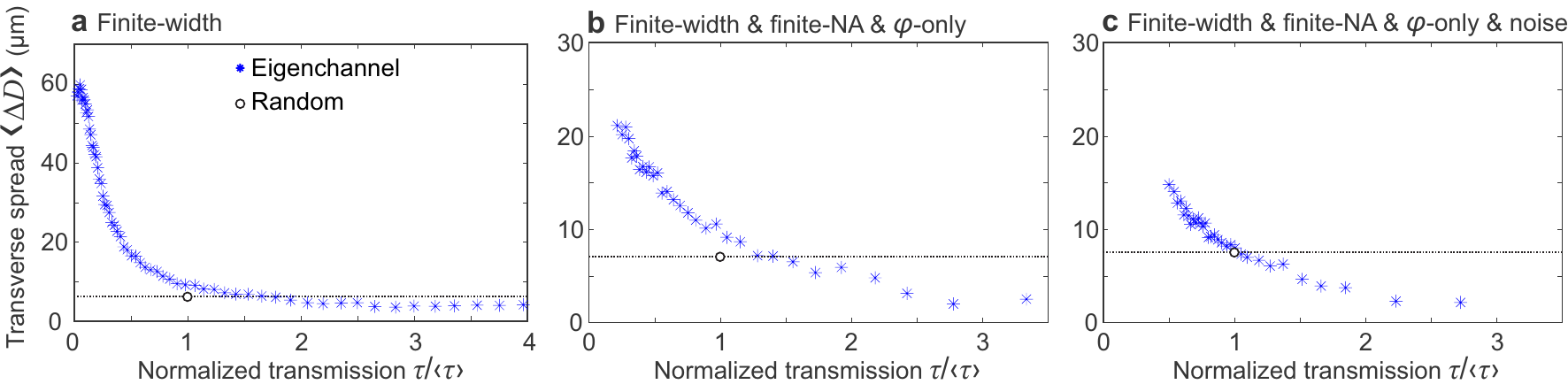}
		\caption{\textbf{Transverse spread of transmission eigenchannels.} Numerically calculated transverse spread length $\Delta D = D_\mathrm{out} - D_\mathrm{in}$ of transmission eigenchannels versus normalized transmission eigenvalue $\tau/\langle\tau\rangle$. \textbf{a}, Only finite-width illumination is considered in the simulation, \textbf{b}, Including finite-NA and phase-only modulation. \textbf{c}, Adding Gaussian random noise to the transmission matrix. In all cases, as $\tau/\langle\tau\rangle$ increases,  $\Delta D$ decreases, indicating high-transmission channels spread less and low-transmission channels spread more than random incident wavefronts (black open circle and dashed line). The slab parameters are the same as those in Fig.~5 of the main text. Each data point represents an average over 50 disorder realizations.}
		\label{figure10}
		\end{figure*}
        
        \begin{figure*}[ht]
		\centering
		\includegraphics[width=\linewidth]{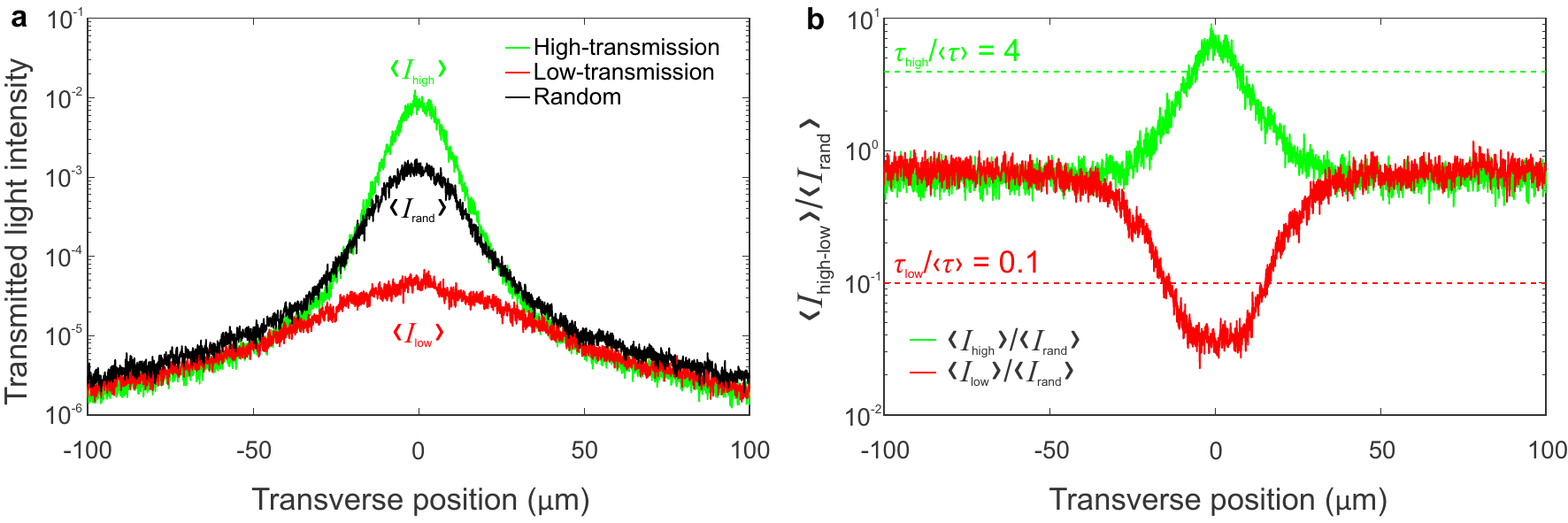}
		\caption{\textbf{Transmitted light intensity profile of a diffusive slab with finite-width illumination.}  \textbf{a}, Spatial profile of light intensity at the output surface of the slab for a high-transmission channel $\left(\tau_\text{high}/\langle\tau\rangle = 4\right)$, a low-transmission channel $\left(\tau_\text{low}/\langle\tau\rangle = 0.1\right)$, and a random incident wavefront, plotted by green, red, and black lines respectively. Each curve is an ensemble average over 50 disorder realizations. The slab parameters are the same as those in Fig.~\ref{figure10}. \textbf{b}, Ratio of the transmitted light intensity profile for high or low-transmission channel to that of random incident wavefront, $\langle I_{\rm high}\rangle / \langle I_{\rm rand}\rangle$ (green line), $\langle I_{\rm low}\rangle/ \langle I_{\rm rand}\rangle$ (red line), illustrating the transmitted light intensity is enhanced or suppressed more at the center than at the tails of the output beam.}
		\label{figure11}
		\end{figure*}

		We excite a single channel and calculate its field distribution across the slab. The input and output widths are obtained from the participation number of field intensity distribution on the front and back surfaces of the slab. We repeat this calculation for slabs of different width $k_0W =$ 100, 250, 500, 1000, 2000, 3000, 4500, 6000, while fixing the thickness $k_0L$, the average refractive index $n_0 = n_1 = n_2$, and the strength of disorder $\sigma$. The channel widths are averaged over all open channels with transmission eigenvalue $\tau_n\geq 1/\text{e}$. As shown in Fig.~1b of the main text, the open channel width $k_0 D$ increases as $W$ increases, eventually saturates to a constant value, which can be extracted from a two-parameter fit,
	\begin{equation}
	\label{fit}
	\langle k_0D\rangle = \left(a_0 + \frac{a_1}{k_0W}\right)^{-1}.
	\end{equation}
	The asymptotic open channel width $k_0 D_\infty = 1/a_0$ is obtained in the limit $W\rightarrow \infty$. 

	We apply the above procedure to slabs of different thickness $k_0L = 50, 100, 150, 200, 250, 300$ to find the scaling of $D_\infty$ with $L$, while fixing $n_0$ and $\sigma$. To find the scaling of $D_\infty$ with $l_t$, the disorder strength in the slab is varied as $\sigma =$ 0.6, 0.65, 0.75, 0.9, 1.1, 1.45. The corresponding normalized transport mean free paths $kl_t$ are given in table~\ref{table}. For each set of parameters $(k_0L, \sigma, k_0W)$, we simulate 10 different realizations of disorder to obtain the ensemble-averaged values. In addition to $n_0 = n_1 = n_2 = 1.5$, we also set $n_0 = n_1 = n_2 = 1$, then vary the thickness $k_0L$ (as listed above) and $\sigma =$ 0.6, 1.1 ($kl_t = 17.4, 6.1$). To check whether the transverse boundary conditions affect the channel widths, we repeat the simulations with periodic boundary conditions, and obtain the same asymptotic open channel widths shown in Fig.~3 of the main text. 

	Finally, we examine whether the scaling of $D_\infty$ depends on the way of averaging. Instead of obtaining $D_\infty$ from $\langle D \rangle$ in the limit of $W \rightarrow \infty$, we compute $\langle 1/D\rangle$ and extract $D_\infty$ from its inverse in $W \rightarrow \infty$ limit. The same scaling of $D_\infty\propto kl_tL$ is obtained, as shown in Fig.~\ref{figure9}, only the prefactor of 0.54 is slightly different from that of 0.68 in Fig. 3c of the main text.
   
  	\subsubsection{Incomplete control}
    
	Experimentally only a small region of a wide slab is illuminated, and partial transmission matrix is measured. We numerically investigate the effects of finite-width illumination, finite numerical aperture (NA), phase-only modulation, and noise on the spatial profiles of transmission eigenchannels.  
	
	We first calculate the complete transmission matrices of 2D slabs for 50 disorder realizations. The slab parameters, given in the caption of Fig.~5 of the main text, are chosen to be close to those of the ZnO nanoparticle layer in our experiment. The slab ($n_0 = 1.4$) is sandwiched between air ($n_1 = 1.0$) and glass ($n_2 = 1.5$). Periodic boundary conditions are applied to the transverse boundaries. The number of input modes (from the air) is $N_1 = 1999 \approx 2n_1W/\lambda$, and the number of output modes (to the glass) $N_2 = 3239$. To model the binning of SLM pixels into macropixels and the limited field of view of detection optics, we group the input and output modes in $k$-space. The number of input modes in one group, $m_1$, is chosen such that the corresponding illumination width on the front surface of the slab, as measured by the participation number of random inputs, is similar to that in the experiment. The number of output modes in a group, $m_2$, sets the size of detection region, which is similar to the field of view in our experiment (180 \textmu m in diameter). Such grouping effectively reduces the number of degrees of freedom to $M_1 = 62$ at the input and and $M_2 = 1079$ at the output. The spatial profile of each transmission eigenchannel is then calculated. The input and output widths are plotted in Fig.~5a of the main text, and the transverse spread is plotted as a function of the normalized transmission eigenvalue in Fig.~\ref{figure10}a. Low-transmission channels spread significantly more than random incident wavefronts, while high-transmission channels spread slightly less than random wavefronts with finite illumination width.

    In order to account for finite NA in the experiment, we take only $M_1 = 32$ input degrees of freedom and $M_2 = 234$ output degrees of freedom for the $k$-space transmission matrix. The ratio $M_1/M_2 = 0.136$ is chosen to match that in the experiment, even though the experimental values of  $M_1$ and $M_2$ are much larger. Additionally, we remove amplitude modulation from the input eigenvectors to simulate phase-only modulation in our experiment. As shown in Fig.~\ref{figure10}b, the range for transverse spread of transmission eigenchannels is reduced, but the trend is similar to that in Fig.~\ref{figure10}a. 
    
    Finally we use random Gaussian noise to model experimental errors in a transmission matrix measurement. We simulate the common-path interferometric measurement using the partial transmission matrix, adding random Gaussian numbers to the intensity values in transmission to model measurement noise. Such random Gaussian noise results in phase estimation errors in the ``measured" transmission matrix, $\tilde{t} + \delta\tilde{t}$, which deviates from the actual matrix $\tilde{t}$ by $\delta\tilde{t}$ due to noise. We calculate the transmission eigenvectors of this partial transmission matrix, remove their amplitude modulations, and then calculate their transmissions and spatial profiles using the actual matrix $\tilde{t}$. The error in the measurement of the transmission matrix cause a further reduction in the range of transmission eigenvalues and the lateral spread of the eigenchannels as Fig.~\ref{figure10}c shows.
    
	The above numerical simulation results illustrate that the finite illumination width has the predominant effect on the spatial profiles of transmission eigenchannels, making their behavior qualitatively different from the eigenchannels of the complete transmission matrix for a wide slab. For comparison, we plot in Fig.~\ref{figure11}a the ensemble-averaged intensity distribution on the output surface of the slab for a high-transmission eigenchannel $\langle I_{\rm high}\rangle$ and a low-transmission eigenchannel $\langle I_{\rm low}\rangle$, in comparison to a random incident wavefront $\langle I_{\rm rand\rangle}$. Their ratios $\langle I_{\rm high}\rangle/ \langle I_{\rm rand}\rangle$ and $\langle I_{\rm low}\rangle/ \langle I_{\rm rand}\rangle$, plotted in Fig.~\ref{figure11}b, reveal that the enhancement of transmitted light intensity is higher at the center of the output beam for the high-transmission channel, leading to an effective reduction of the width (characterized by the participation number) of the output beam. For the low-transmission channel, the suppression of transmitted light intensity is also stronger at the beam center, resulting in an increase of output beam-width. Such behavior is attributed to the fact that multipath interference effects are enhanced near the center of illumination region due to higher probability of scattering path crossing compared to the edges.

% Bibliography
\bibliography{transverse_localization}

\end{document}